\documentstyle[prl,aps]{revtex}
\begin{document}

\gdef\journal#1, #2, #3, 1#4#5#6{               
    {\sl #1~}{\bf #2}, #3 (1#4#5#6)}            
\def\prb{\journal Phys. Rev. B, }
\def\prl{\journal Phys. Rev. Lett., }
\def\jltp{\journal J. Low Temp. Phys., }

\def\lan{\langle}
\def\ran{\rangle}
\def\nono{\nonumber}
\def\d{\delta}
\def\D{\Delta}
\def\t{{\theta}}
\def\a{{\alpha}}
\def\b{{\beta}}
\def\beqar{\begin{eqnarray}}
\def\eeqar{\end{eqnarray}}
\def\e{{\epsilon}}
\def\beq{\begin{equation}}
\def\eeq{\end{equation}}
\def\s{{\sigma}}
\def\L{{\Lambda}}
\def\va{{\vec a}}
\def\vr{{\vec r}}
\def\vk{{\vec k}}
\def\ve{{\vec e}}
\def\vmu{{\vec \mu}}
\def\vm{{\vec m}}
\def\vex{{\vec e_x}}
\def\vey{{\vec e_y}}
\def\vG{{\vec G}}
\def\vQ{{\vec Q}}
\def\vq{{\vec q}}
\def\vS{{\vec S}}
\def\hO{{\hat \Omega}}
\def\ad{{a^\dagger}}
\def\bd{{b^\dagger}}
\def\cd{{c^\dagger}}
\def\g{{\gamma}}
\def\o{{\omega}}
\def\he{{He$^4$}}
\def\sm{{S_{-}}}
\def\sp{{S_{+}}}
\def\tS{{\tilde{S}}}
\def\qtr{{1\over4}}
\def\tend{{\rightarrow}}
\def\kag{{Kagom\'e\ }}
\def\roott{{$\sqrt{3}\times\sqrt{3}$}\ }

\def\frac#1#2{{\textstyle{#1\over #2}}}
\def\ssrm{\scriptscriptstyle\rm}
\def\yd{^\dagger}
\def\nd{^{\vphantom\dagger}}
\def\lra{\leftrightarrow}
\def\cH{{\cal H}}
\def\cA{{\cal A}}
\def\ij{{\lan ij\ran}}
\def\zhat{{\hat z}}
\def\xhat{{\hat x}}
\def\yhat{{\hat y}}
\def\pz{\partial}
\def\etal{{\it et al\/}\ }
\def\ie{{\it i.e.\/}\ }
\def\eg{{\it e.g.\/}\ }
\def\etc{{\it etc.\/}\ }
\def\Neel{N{\'e}el solid\ }
\def\half{\frac{1}{2}}
\def\third{\frac{1}{3}}
\def\twothirds{\frac{2}{3}}
\def\fourth{\frac{1}{4}}
\def\thalf{\frac{3}{2}}
\def\rtot{\frac{\sqrt{3}}{2}}
\def\ta{\theta_{\ssrm A}}
\def\tb{\theta_{\ssrm B}}
\def\tc{\theta_{\ssrm C}}
\def\fa{\phi_{\ssrm A}}
\def\fb{\phi_{\ssrm B}}
\def\fc{\phi_{\ssrm C}}
\def\eo{e_{\ssrm MF}}
\def\Kag{{\ssrm Kag}}
\def\Tri{{\ssrm Tri}}
\def\SF{{\ssrm SF}}
\def\SS{{\ssrm SS}}
\def\NS{{\ssrm NS}}
\def\MI{{\ssrm MI}}
\def\MF{{\ssrm MF}}
\def\SW{{\ssrm SW}}
\def\rmi{{\rm i}}
\def\rmc{{\rm c}}
\def\vM{{\vec M}}
\def\vR{{\vec R}}
\def\vRn{{{\vec R}\nu}}
\def\cO{{\cal O}}
\def\nup{{\nu'}}
\def\nnp{{\nu\nu'}}
\def\vdelta{{\vec\delta}}
\def\hef{${}^4$He\ }

\title{Superfluids and Supersolids on Frustrated 2D Lattices}
\author{Ganpathy Murthy}
\address{Department of Physics and Astronomy, The Johns Hopkins
  University\break 
(Permanent address: Department of Physics, Boston University)}
\author{Daniel Arovas}
\address{Physics Department 0319, University of California at San Diego,
La Jolla CA 92093-0319}
\author{Assa Auerbach}
\address{Department of Physics, Technion IIT, Haifa, Israel}
\maketitle
\begin{abstract}
We study the ground state of hard-core bosons with nearest-neighbor
hopping and nearest-neighbor interactions  on the triangular and
\kag lattices by mapping to a system of spins ($S=\half$), which we
analyze using spin-wave theory. We find that the both lattices display
superfluid and supersolid (a coexistence of superfluid and solid) order
as the parameters and filling are varied. Quantum fluctuations seem
large enough in the \kag system to raise the interesting possibility
of a disordered ground state. 

\end{abstract}

\section*{Introduction}
\label{sec:intro}

A supersolid is a state of matter which simultaneously exhibits both
solid and superfluid properties.  That is to say, it displays both
long-ranged positional order as well as finite superfluid density and,
na{\"\i}vely, off-diagonal long-ranged order (ODLRO).  The intriguing
suggestion by Andreev and Lifshitz \cite{andreev} that vacancies in
solid ${}^4$He might Bose condense in the vicinity of the melting line
has, to our knowledge, never been experimentally verified
\cite{meisel,lengua}.  Nonetheless, a sizeable literature has developed
on the theoretical properties of supersolids
\cite{chester,leggett,bruder,vo1,vo2,erds,jfhm,nsfb1,nsfb2,akgz,ucdavis,pomeau,other}. In two dimensions, the physics of Josephson
Junction Arrays\cite{doniach} has also stimulated the theoretical study of
supersolids. Most of the work is based on the contributions 
 of Matsuda and Tsuneto \cite{matsuda} and of Liu
and Fisher \cite{liu-fisher}, who established some key concepts in the
theory of lattice-based supersolid models.  Of central importance is the
mapping between a hard-core lattice Bose gas and a spin-$\half$ quantum
magnet:
\begin{equation}
a\yd_i\lra S^+_i\qquad a\nd_i\lra S^-_i\qquad n_i=a\yd_i a\nd_i\lra S^z_i-\half\ .
\label{mapping}
\end{equation}
Thus, an occupied site is represented by an up spin, while an empty site is
represented by a down spin.  An interacting hard core lattice Bose gas with
nearest neighbor hopping $t$, nearest neighbor repulsion $V$, and chemical potential
$\mu$ is thereby equivalent to the anisotropic $S=\half$ Heisenberg model
\begin{equation}
\cH=\sum_\ij \left(J_\parallel\, S^z_i S^z_j + J_\perp\,(S^x_i S^x_j +
 S^y_i S^y_j)\right)-H\sum_i S^z_i
\label{Heis}
\end{equation}
where $J_\parallel=V$ is an antiferromagnetic longitudinal exchange,
$J_\perp=-2t$ is a ferromagnetic transverse exchange, and $H=\mu-\half
zV$ is an external magnetic field ($z$ is the lattice coordination
number).  The spin model exhibits a global U(1) symmetry with respect to
rotations about the $\zhat$ axis, which of course means total particle
number conservation in the boson language. The boson condensate order
parameter is related to the transverse magnetization density via $\lan
a\yd_i\ran=\lan S^+_i\ran$, while the boson compressibility $K=\pz
n/\pz\mu$ is the magnetic susceptibility $\chi=\pz M_z/\pz H$.  Liu and
Fisher identified four phases of interest: (i) a normal fluid, in which
the magnetization is uniform and in the $\zhat$ direction, (ii) a normal
solid, in which the magnetization lies along $\zhat$ yet is spatially
modulated at some wavevector $\vk$, (iii) a superfluid, in which the
magnetization is uniform and has a component which lies in the $x$-$y$
plane, and (iv) a supersolid, in which there simultaneously exists a
nonzero transverse component to the magnetization $\vM_\perp$, as well
as a spatial modulation of the longitudinal magnetization $M_z$.  (An
incompressible normal fluid is also called a Mott insulator.)

If one relaxes the hard core constraint in favor of a finite on-site repulsion $U$,
one obtains the Bose Hubbard model
\begin{equation}
\cH=-t\sum_\ij(a\yd_i a\nd_j + a\yd_j a\nd_i) -\mu\sum_i n_i
+\half U\sum_i n_i(n_i-1) + V\sum_\ij n_i n_j\ .
\end{equation}
This model has been extensively studied since the seminal work of
Fisher, Weichman, Grinstein, and Fisher\cite{fwgf}, who considered the
model with $V=0$ in the context of a superconductor-insulator
transition. A study of this model in the presence of disorder has led to
an understanding of the Bose glass\cite{wallin}. On a two-dimensional
square lattice, and for $V\ne0$, the model was studied at $T=0$, for
both finite and infinite $U$, by Scalettar \etal \cite{ucdavis}, Bruder
\etal\cite{bruder}, and van Otterlo \etal\cite{vo1,vo2} using mean field
theory and quantum Monte Carlo techniques.  To summarize their results,
no supersolid phase is observed at half filling ($\lan n\ran=\half$),
where a first order transition occurs as a function of $V$ between
superfluid (small $V$) and a N\'eel solid (large $V$, $\vk=(\pi,\pi)$).
(Large next-nearest neighbor repulsion $V'$ stabilizes a striped phase,
the colinear solid.)  Away from half filling, there is no normal solid
phase, and the transition is instead from superfluid to supersolid.  The
supersolid phase exhibits both a peak in the static structure factor
$S(\vk)$ at the N\'eel vector (and properly proportional to the lattice
volume), and a nonzero value of the superfluid density $\rho_{\rm s}$.
Again, next nearest neighbor $V'$ can stabilize a striped supersolid
phase with anisotropic $\rho_{\rm s}$. One can also obtain Mott
insulating phases with fractional filling in the presence of
next nearest neighbor interactions. 

In this paper, we will investigate the properties of the model in eq.
\ref{Heis} on frustrated two-dimensional lattices. We are motivated by
the fascinating interplay between frustration, quantum fluctuations,
order, and disorder which has been seen in quantum magnetism.

  Frustration enhances the effects of quantum
fluctuations.  Indeed, as early as 1973, Fazekas and Anderson \cite{anderson,fazekas-pwa}
raised the possibility that for such systems, quantum fluctuations
might destroy long-ranged antiferromagnetic order even at zero temperature.  
In many cases, frustration leads to an infinite degeneracy at the classical
(or mean field) level not associated with any continuous symmetry of the Hamiltonian
itself.  In these cases, it is left to quantum (or thermal) fluctuations to
lift this degeneracy and select a unique ground state \cite{shender1,henley1},
sometimes with long-ranged order.  Our models exhibit both a depletion (but not
unambiguous destruction) of order due to quantum fluctuations, as well as the
phenomenon of ``order by disorder''.

In our work, we will choose the units of energy to be $J_\parallel$,
writing $\Delta\equiv t/V=J_\perp/ 2 J_\parallel$, and $h\equiv H/J_\perp$.
We will be following closely the analysis of the anisotropic triangular
lattice antiferromagnet by Kleine, M\"uller-Hartmann, Frahm, and Fazekas
(KMFF)\cite{kmff}, who performed a mean field ($S=\infty$ limit) and spin wave theory
(order $1/S$ corrections to mean field) analysis.
Comtemporaneously with KMFF, Chubukov and Golosov\cite{chub-gol} derived
the spin-wave expansion for an isotropic Heisenberg antiferromagnet in a magnetic
field, while Sheng and Henley\cite{sheng} obtained the spin-wave theory for the
anisotropic antiferromagnet in the absence of a field. 

The mean field phase diagram is shown in figure 1 (both the triangular
and \kag lattices have the same mean field phase diagram up to a
rescaling of $h$). Notice that the supersolid phase appears in a broad
region of $\D$ and filling. The reason the supersolid is so robust is
that the lattice frustrates a full condensation into a solid.
Generically, frustrated lattices might be good places to look for this
phase.

Let us briefly concentrate on $h=0$ before describing the entire phase
diagram. We will be assuming a three sublattice structure throughout. 
The mean field state is then described by three polar and three azimuthal
angles: $(\ta,\tb,\tc,\fa,\fb,\fc)$, and is invariant under uniform rotation
of the azimuths.

Due to the ferromagnetic coupling in the $x-y$ spin directions the
mean field solution is always coplanar.  Just as in KMFF, there is a
one-parameter family of degenerate mean field solutions in the
zero-field case (originally found by Miyashita and
Kawamura\cite{miya-kawa}).  The A sublattice polar angle $\ta$
may be chosen as the free parameter; spin wave theory (SWT) is
necessary to lift the degeneracy and uncover the true ground state.
Figure 2 shows the ground state energy in SWT as a function of $\ta$ for
the triangular lattice at $\D=0.25$.  Using SWT we also compute the
fluctuations of the spins, and the consequent quantum-corrected
magnetization and the solid and ODLRO order parameters. Figure 3
illustrates these quantities in mean field and to leading order in SWT
(where $S$ has been set equal to $\half$) 
as a function of $\D$ for the triangular lattice. It is clear that the
quantum corrected $S^z$ is very close to zero for all $\D$,
reflecting the fact that at $h=0$ the lattice is half-filled. Two
sublattices acquire large corrections due to quantum fluctuations (even
in the Ising limit $\D\rightarrow0$), while the third has only small
quantum corrections. This is very similar to the fully antiferromagnetic
case studied by KMFF.   Therefore, even at $S=\half$ the solid order
survives. The off-diagonal order parameter $S^x$ is reduced
in magnitude by quantum corrections, but goes to zero only as $\D\tend 0$.
Supersolid order survives quantum fluctuations for the
triangular lattice, at least in this order of SWT.

Let us now consider the \kag lattice. There is a qualitative difference
between the antiferro-ferromagnetic case considered here, $\D\ge 0$, and
the fully antiferromagnetic case $\D\le 0$, which has been exhaustively
explored for the Heisenberg limit $\D=-1$ (for a partial set of
references, see \cite{elser,zeng1,zeng2,marston,harris,subir-spn,chandra1,chandra2}).
For the fully antiferromagnetic case there are local motions of the spins that
move the system on the degneracy submanifold, leading to a much larger
ground state degeneracy for the \kag lattice than for the triangular
lattice.  However, for $\D\ge 0$, the ferromagnetic transverse interaction
eliminates the possibility of these local motions,
resulting in a ground state degeneracy parametrized only by $\ta$, just
as in the triangular lattice.

We carried out SWT for the two long-range ordered configurations shown
in figure 4 -- a three sublattice ``$\vq=0$'' state, and a nine
sublattice \roott structure\cite{harris,subir-spn}, respectively.  These
states have the same mean field ground state energy.  It will turn out
that two of the three sublattices have the same spin orientation in the
ground state for both lattices. With this proviso, note that the $\vq=0$
structure has one two-fold axis and a mirror plane (point group
$C_{2v}$), while the \roott structure has a six-fold axis and a mirror
plane (point group $C_{6v}$).  Once again SWT selects the true ground
state. When quantum fluctuations are accounted for, we find that the
\roott structure always has lower energy than the $\vq=0$ structure, to
the numerical accuracy of our calculations.  More importantly, the
fluctuations of the spins on six of the nine sublattices diverge in the
limit $h\tend 0$, as shown in figure 5. This divergence is the
consequence of a flat (dispersionless) mode at zero energy as $h\tend0$.
Higher order terms in the spin wave expansion will lift this mode and
remove the divergence \cite{shender1,stevek}.  However, the fluctuations
remain large for $S=\half$, as we estimate in section III.  This
indicates that quantum flucuations may be strong enough to wash out any
order, including ODLRO, on two of the sublattices, which raises the
intriguing possibility of a partially disordered ground state at $T=0$.
It must be emphasized that we have not demonstrated that this is so: the
fluctuations could be correlated between different sites, there could be
long-range order with a larger unit cell, a condensed array of vortices,
\etc\ \ This problem merits further study, with \eg quantum Monte Carlo
methods.

Let us now turn to a fuller description of figure 1, where four types of
mean field states are present (solid lines indicate first order transitions).
We adopt the nomenclature of ref. \cite{ucdavis} (see fig. 12 of this reference
for comparison).  At high fields $h$ the system is in a Mott
phase -- incompressible and fully polarized.  As the field is lowered, for
any $\Delta >0$, the system enters a compressible superfluid
phase (SF), with $\ta=\tb=\tc>0$.  A first order transition from the superfluid
to an incompressible N{\'e}el solid (NS) at filling fraction $\frac{2}{3}$
(magnetization per site $M_z=\pm\frac{1}{3} S$) occurs for $\Delta < \half$.
Finally, the supersolid phase (SS) exists for $\Delta < \half$ between the
two symmetry-related \Neel lobes.  There is a tricritical point at
$(\Delta^*,h^*)=(\half,3)$.  Quantum fluctuations will modify these mean field
phase boundaries.  Since the superfluid state, we have found, benefits the
most from spin wave energy corrections, it will encroach on its
neighbors as $S$ decreases from $\infty$.  Increasing $h$ tends to suppress
quantum fluctuations.

We will present each of the above results in more detail in the rest of
this paper. Section I concentrates on the mean field theory and the
mean field phase diagram. Section II describes the selection of the true
ground state by quantum fluctuations at $h=0$, and the form of the
spin-wave excitations for arbitrary $h$. Section III presents the
suppression of order by quantum fluctuations. We end with our
conclusions, connections to experimental work, and open questions in
Section IV.

\bigskip
\section{Mean Field Theory}
\bigskip

The mean field limit is obtained by setting $S=\infty$. Formally, we
first generalize the model from $S=\half$ to a model with a spin-$S$ at
each site. We then represent each spin as a classical vector of
magnitude $S$, $S^{\a}=S\Omega^{\a}$, where $\hO$ is a unit vector in three
dimensions.  We rescale the magnetic field by $S$ and write
\beq
\cH_\MF/S^2=\sum_\ij\Omega^z_i\Omega^z_j-\Delta \sum_\ij (\Omega^x_i\Omega^x_j
+\Omega^y_i\Omega^y_j)-h\sum_i\Omega^z_i\ .
\label{HMF}
\eeq
In the mean field solution, all spins lie in the $x$-$z$ plane.  Furthermore,
it turns out that the triangular and both \kag structures have the same mean
field energy, to within a constant factor (up to a separate rescaling of $h$
in the case of the \kag structures), so we will consider all three cases
simultaneously.  Note that coplanar states have been selected at the mean field
level for the \kag lattice, in contrast to the fully antiferromagnetic case.

\bigskip
\subsection{Zero Field}
\medskip

Consider first mean field ground states on the triangular and \kag lattices
that have a three-sublattice structure.  The three
specific cases are the usual sublattice structure on the triangular
lattice, the $\vq=0$ structure on the \kag lattice, and the \roott
structure on the \kag lattice.  The nine sublattices of the \roott structure
are organized into three groups (A,B,C) of three, so that an A site has two B
and two C neighbors.  Thus, the energy per site, in units of $S^2$, is
\beqar
\eo\equiv E_\MF/NS^2&=(\cos\ta\cos\tb+\cos\tb\cos\tc+\cos\tc\cos\ta)\cr
&-\D(\sin\ta\sin\tb+\sin\tb\sin\tc+\sin\tc\sin\ta)
\eeqar
on the triangular lattice and $z_\Kag/z_\Tri=\frac{2}{3}$ this value
on the \kag lattice.

Miyashita and Kawamura\cite{miya-kawa} have shown that there is a
one-parameter family of degenerate ground states for this classical
problem for arbitrary $\D$, which does not seem to be related to any
obvious symmetry of the model. Following KMFF, and writing
$\b_i=\t_i-(\ta+\tb+\tc)$, and defining the two-dimensional vectors
$\vmu_i=(\sin\b_i,\cos\b_i)$, we can write the mean field energy in terms
of the two-component vector $\vmu\equiv\vmu_{\ssrm A}+\vmu_{\ssrm B}+
\vmu_{\ssrm C}$:
\beq
\eo=\fourth(1-\Delta)(\vmu^2-3)+\half(1+\Delta)\mu_y\ .
\eeq
Therefore $\eo$, while nominally depending on the three angles
$\t_i$, actually depends only on two combinations of them, leaving one
parameter free.  We can then parameterize the degenerate ground states
by $\ta$, by defining $\tb=\e-\d$, and $\tc=\e+\d$, where
\beqar
\tan\e&=&-{\tan\ta\over\D}\cr
\cos\e&=&{-\D\cos\ta\over\sqrt{1-(1-\D^2)\cos^2\ta}}\cr
\cos\d&=&{\D\over{(1-\D)\sqrt{1-(1-\D^2)\cos^2\ta}}}
\label{angles}
\eeqar
It is easy to verify that $\ta$ can lie in the range
$\cos^{-1}\left({1\over1-\D}\sqrt{{1-2\D\over1-\D^2}}\right)
\leq\ta\leq\half\pi$. 

As $\D$ increases from zero, the range of possible $\ta$ is compressed, and the
differences $|\tb-\ta|$, $|\tc-\ta|$ shrink, until at $\Delta=\half$ the ground
state is colinear with $\ta=\tb=\tc=\half\pi$ -- a featureless superfluid.
For $\Delta<\half$, in the case of zero field, the true ground state will be
selected by quantum fluctuations.  Minimizing $\eo$ gives
$\mu_x=(\Delta+1)/(\Delta-1)$, $\mu_y=0$, and
\beq
\eo=-\left({1-\Delta+\Delta^2\over1-\Delta}\right)\ .
\eeq

\bigskip
\subsection{Non-zero Field}
\medskip

Turning on a field $h$ adds an energy 
\beq
\Delta\eo= -\third h(\cos\ta+\cos\tb+\cos\tc)
\eeq
per site and lifts the degeneracy described in the previous subsection,
producing a unique mean field ground state (unlike in the Heisenberg
antiferromagnet\cite{chub-gol}).  We find that minimization generally leads
to a state where (without loss of generality) $\ta=\tb\neq\tc$ in the
supersolid phase.  The results are plotted in figure 6.  A rescaling of the
field for the \kag lattice ($h_\Kag=\twothirds h_\Tri$) makes the entire mean
field phase diagram identical, and we have therefore shown only the
triangular lattice results.

The phase diagram has already been shown in figure 1.  Let us keep at a
particular value of $\Delta$ and turn up the field $h$.  At zero field there are
two regimes, the superfluid with no solid order for $\Delta>\half$, and the
supersolid with both solid and ODLRO ($T=0$) for $\Delta<\half$. 
For $\D>\half$, the ground state remains a uniform superfluid, though $M_z$
becomes nonzero, as $h$ is increased.  The spins cant at an angle
$\theta=\cos^{-1}(h/6(1+\Delta))$.  The energy in this phase is
\beq
\eo^{[\SF]}=-3\D-{h^2\over12(1+\D)}\ .
\eeq

Eventually, for $h>6(1+\D)$, every site has the maximum possible $S^z$,
and the system is in the MI phase, with each site fully occupied with one boson.
Borrowing from spin-wave results derived in section II, the  linear instability of
the SF phase occurs at  
\beq
h_{\rmi 1}=6\sqrt{(1+\D)(1-2\D)}
\eeq

Next focus on a specific $\Delta<\half$, and increase the field $h$ from
zero.  At zero field, the one-parameter degeneracy of the mean field
ground states has to be lifted by quantum fluctuations, which select a
particular $\ta$.  However, it turns out that we can recover this $\ta$
by considering nonzero $h$ and taking the limit $h\tend 0$, which gives
\beq \cos^2 \ta = {1-2\Delta \over 1-\Delta^2}\ .  \eeq It seems
surprising that the ground state selected by quantum fluctuations can be
predicted by an entirely classical calculation. A plausible (though
non-rigorous) argument will be provided for this in terms of spin-wave
theory in the next section.

As $h$ is increased from zero, $\ta$ and $\tc$ change (recall $\ta=\tb$
throughout).  At a certain critical field $h_{\rmc 1}(\Delta)$, $\vS_C$ points
exactly along the $-\zhat$ direction ($\tc=\pi$), while $\vS_{A,B}$ point
along the $\zhat$ direction ($\ta=\tb=0$).  This critical field $h_{\rmc 1}$
can be analytically determined by the following consideration: the point
$\ta=0,\ \tc=\pi$ is always a stationary point of the mean field energy.
However, for $h<h_{\rmc 1}$ it is a saddle point, while for $h>h_{\rmc 1}$
it is a minimum. Therefore the second derivative matrix of $\eo(\ta,\tc)$
should have a zero eigenvalue at $h=h_{\rmc 1}$.  Setting the determinant
to zero gives
\beq
h_{\rmc 1}(\Delta)=\frac{3}{2}(2+\Delta-\sqrt{4-4\Delta-7\Delta^2})
\eeq

For $h<h_{\rmc 1}(\Delta)$, we are in the supersolid phase.  Just above
$h_{\rmc 1}$, however, the system is exactly $\twothirds$ filled, and
its magnetic susceptibility is zero (the compressibility of the
corrresponding boson system is zero). There is no superfluid order and
the system is again a Mott insulator, the \Neel.  It will be seen in
section III that the exact filling of $\twothirds$ survives quantum
fluctuations.  Since the superfluid order goes continuously to zero
below the transition, it is clear that this transition is second-order
within in mean field theory.  The energy of the \Neel phase is

\beq 
\eo^{[\NS]}=-1-{h\over3}
\eeq
One can furthermore determine that the \Neel phase is linearly stable for
$\thalf(2+\D-\sqrt{4-4\D-7\D^2})\le h\le {3\over2}(2+\D+\sqrt{4-4\D-7\D^2})$.

Further increasing $h$, we find a critical field
\beq
h_{\rmc 2}(\Delta)=2(1+\Delta)+4\sqrt{(1+\Delta)(1-2\Delta)}
\eeq
beyond which the canted superfluid becomes energetically favored.
The transition is first-order since it is far from any linear
instabilities.  Finally, a second order line at $h>h_{\rmc 3}(\Delta)=6(1+\Delta)$,
signals the boundary between superfluid and fully polarized Mott phases.

When $\Delta=\half$, we can solve analytically to find
\beqar
\cos\ta&=&\third h\cr
\cos\tc&=&-\third h\cr
\eo^{[\SS]}&=&-\thalf-\frac{1}{18} h^2
\eeqar
This is exactly the same as the energy of the SF phase at $\Delta=\half$ (which is
characterized by all the angles satisfying $\cos\t=\frac{1}{9} h$), which marks this
vertical line as a first order line.  Note the tricritical point at
$\Delta=\half,\ h=3$, where two first order lines (with infinite slope) meet
a second-order line (with finite slope).

The situation is completely symmetric with respect to the sign of $h$,
with the \Neel phase now existing at $\third$ filling for $h<0$.  At the
mean field level, the $\Delta$-axis is a first-order transition line up to
$\Delta=\half$, since $S^z$ is discontinuous across it.  However, for
$S=\half$ it apparently becomes continuous, at least for the triangular
lattice.

The phase diagram has some similarities to the classic picture of the
Mott lobes surrounded by superfluid described by Fisher, Weichman,
Grinstein, and Fisher (FWGF) \cite{fwgf}.  However, there are important
differences.  FWGF considered local Hubbard repulsion, whereas our extended
Bose Hubbard model we consider affords the possibility of incompressible Mott phases
at fillings 0, $\third$, $\twothirds$, and $1$.  The \Neel phase found by
Scalettar \etal, for example, exists at filling $\half$.  Fractional
fillings have also been observed in the square lattice with frustrating
longer range interactions in ref. \cite{bruder}.  Also, the transitions 
>from the fractional filling MI phases to the (canted) superfluid are
first order.  Finally, and most notably, the entire region between the
two fractional Mott lobes is taken over by the supersolid, and the
supersolid gives way to the superfluid only beyond a hopping
$t>\half V$.

\bigskip
\section{Spin-Wave Theory}
\bigskip

We now develop the spin wave theory (SWT) for this problem.  If $h\neq 0$, there is
a unique ground state (up to permutations of the sublattices).  When $h=0$,
the ground state manifold is parameterized by $\ta$, and the other angles,
$\tb$ and $\tc$ (in general not equal), can be determined from $\ta$ and $\Delta$
using eq. \ref{angles}.  We implement SWT in the usual way: by first
performing local rotations of the spins so that the mean field
directions point along the local $z$-axis.  Since all the spins are
assumed to lie on the $x$-$z$ plane, we can do this by a rotation
about the $y$-axis.  Labelling the local frame spins with a tilde, we
have
\beqar
S^x_{\vR\nu}&=&\cos\theta_\nu \tS^x_{\vR\nu} + \sin\theta_\nu \tS^z_{\vR\nu}\cr
S^y_{\vR\nu}&=&\tS^y_{\vR\nu}\cr
S^z_{\vR\nu}&=&-\sin\theta_\nu \tS^x_{\vR\nu} + \cos\theta_\nu \tS^z_{\vR\nu}
\eeqar 
where the subscript $\vR$ labels a Bravais lattice site, $\nu$ a basis element,
and $\theta_\nu$ is $\theta_{A,B,C}$ depending on the mean field orientation
of the $\nu$ sublattice.  The triangular lattice
and the $\vq=0$ structure on the \kag lattice have three sublattices while
the \roott structure on the \kag lattice has nine.

We now describe the spin operators in terms of Holstein-Primakoff bosons
\beqar
\tS^+_\vRn &=&\psi\yd_\vRn\>\sqrt{2S-\psi\yd_\vRn\psi\nd_\vRn}\cr
&=&\sqrt{2S}\>\psi\yd_\vRn+\cO(S^{-1/2})\cr
\tS^-_\vRn &=&\sqrt{2S-\psi\yd_\vRn\psi\nd_\vRn}\>\psi\yd_\vRn\cr
&=&\sqrt{2S}\>\psi\nd_\vRn+\cO(S^{-1/2})\cr
\tS^z_\vRn &=& \psi\yd_\vRn\psi\nd_\vRn-S\ .
\label{HP}
\eeqar
The Hamiltonian (restoring $h$ for generality) is now written in Fourier space as 
\beq
\cH_\SW=E_0 +\half S\sum_\vk \colon\Psi\yd(\vk)\pmatrix{M&N\cr N&M\cr}\Psi\nd(\vk)\colon
+\cO(S^0)
\eeq
(note the normal ordering) where the energy $E_0$ is given by
\beq
E_0=(N/K)S^2\left[\half\sum_\nnp z_\nnp\, X_\nnp - \sum_\nu h_\nu\right] .
\eeq
Here, $N$ is the total number of lattice sites, $K$ is the number of sublattices,
and we define the quantities
\beqar
X_\nnp&=&\cos\theta_\nu\cos\theta_\nup -\Delta\sin\theta_\nu\sin\theta_\nup\cr
Y_\nnp&=&\sin\theta_\nu\sin\theta_\nup -\Delta\cos\theta_\nu\cos\theta_\nup\cr
h_\nu&=& h\cos\theta_\nu
\eeqar
and $z_\nnp$ is the number of $\nu'$ sublattice neighbors each $\nu$ sublattice site has.
The vector $\vk$ lives in the first Brillouin zone of the reciprocal lattice, and
\beq
\Psi\yd(\vk)=(\ \psi\yd_1(\vk), \psi\yd_2(\vk), \ldots, \psi\yd_K(\vk),
\psi\nd_1(-\vk), \psi\nd_2(-\vk), \ldots, \psi\nd_K(-\vk)\ )\ .
\eeq
The matrices $M$ and $N$ have diagonal and off-diagonal elements given by
\beqar
M_{\nu\nu}(\vk)&=&h_\nu-\sum_\nup z_\nnp\, X_\nnp\cr
M_\nnp(\vk)&=&\half(Y_\nnp-\Delta) f_\nnp(\vk)\cr
N_{\nu\nu}(\vk)&=&0\cr
N_\nnp(\vk)&=&\half(Y_\nnp+\Delta) f_\nnp(\vk)
\eeqar
where the function $f_\nnp(\vk)$ is given by the following sum,
\beq
f_\nnp(\vk)={{\sum_\vdelta}}' \exp(i\vk\cdot\vdelta)
\eeq
where the prime on the sum indicates that the sum is over nearest neighbor
vectors connecting a $\nu$ sublattice site to a $\nu'$ sublattice site.

Now we perform a Bogoliubov transformation, which amounts to finding a
rank $2K$ matrix $T$ satisfying $T\yd\Lambda T=\Lambda$, with
\beq
\Lambda=\pmatrix{1_{K\times K}&0_{K\times K}\cr 0_{K\times K}& -1_{K\times K}}\ ,
\eeq
as well as $\Lambda T^{-1}\Lambda\cH_\SW T=\omega$, a nonnegative diagonal matrix with
identical upper left and lower right blocks.  The $\omega_\nu(\vk)$ are the spin wave
frequencies.  The spin-wave correction to the ground state energy is given by
\beq
\Delta E = \half S\sum_\vk\sum_{\nu=1}^K (\omega_\nu(\vk)-M_{\nu\nu}(\vk))\ .
\eeq\

\bigskip
\subsection{Ground State Selection}
\medskip

Figure 2 shows the ground state energy, including spin wave corrections,
for the triangular lattice at $h=0$ as a function of $\ta$.  Since the
classical energy is independent of $\ta$ all the variation comes from
the SW correction. It is clear that there are two possible values of
$\ta$ which minimize the ground state energy, one of them lying at the
edge of the allowed range of $\ta$.  However, the two minima turn out to
be physically identical, and correspond to a relabeling of the sublattices.
We call the minimizing value of $\ta$ which lies away from the edge of its
allowed range $\ta^*$.

The value of $\ta^*$ can be determined by purely classical arguments,
by extremizing the value of $S^z\propto \ta+\tb+\tc$ along the degeneracy
submanifold.  We have checked that this is so by a comparison of the
analytic expression $\ta^*=\cos^{-1}\sqrt{(1-2\Delta)/(1-\Delta^2)}$
with the ground state energy curves obtained from SWT.  We now provide
an argument indicating why this might be the case. 

Our argument will be the following: For a generic $\ta$ there is only
one mode of zero energy at $\vk=0$, whereas if ${\partial
  S_z\over\partial \ta}=0$ there are two zero modes (to spin-wave
order). The ground state energy is the sum of the energies of all the
modes, and does not depend on just the zero modes. However, we think it
plausible that having more modes of zero energy at $\vk=0$ drags down
the energies of all the $\vk$ modes, thus reducing the full ground state
energy. Of course this argument is not rigorous\cite{henley-pc}, and
there may be counterexamples that we are not aware of.

Let us go on to show the first statement about the number of modes of
zero energy. We only sketch the argument here: We treat the issue with
more generality and greater detail in 
the appendix. It is helpful to think in terms of the coherent states path
integral\cite{assa-book} 
\beqar 
{\cal  Z}&=&\int D[\ta,\tb,\tc,\fa,\fb,\fc]\>  e^{-\cA}\cr
\cA&=&\int d\tau\left(\sum_\vr iS\cos\theta(\vr,\tau){\pz\over\pz\tau}
\phi(\vr,\tau)+\cH[\theta,\phi]\right)
\eeqar
where $\cH$ stands for the spin hamiltonian of equation \ref{HMF} written
in terms of $\theta$ and $\phi$.  The first term is the Berry phase
contribution to the path integral, which also makes $S\cos\theta$ the
momentum canonically conjugate to $\phi$.

Let us concentrate on just the $\vk=0$ modes. There are three $\phi$
variables and three conjugate $\t$ variables. We choose to call one of
the $\phi$ variables $\phi_0$ corresponding to an overall rotation of all
the spins around the $z$-axis, and call the remaining $\phi$s $\phi_1$
and $\phi_2$.

Choose any particular ground state labelled by $\ta$. The hamiltonian
for small deviations from the ground state configuration is now given
by
\beq
{\cal H}=\half \t^T M_{\t}\t  + \half \phi^T M_{\phi}\phi
\eeq
where $\t^T=(\delta\ta,\delta\tb,\delta\tc)$ and
$\phi^T=(\phi_0,\phi_1,\phi_2)$. The first row and column of $M_{\phi}$
are zero since $\phi_0$ does not appear in the hamiltonian. 

The general procedure for finding the normal modes is the following:

\begin{itemize}
\item{} Diagonalize the $2\times2$ block of $M_{\phi}$ and rescale the
resulting eigenvectors so that the rescaled $M_{\phi}'$ becomes a unit
matrix in the $2\times2$ block. Call the rescaled $\phi$ variables
$\psi_{\pm}$.  
\item{} Use the Berry's phase terms to identify the canonically
conjugate momenta to $\phi_0$, $\psi_{\pm}$ as $P_0$, $P_{\pm}$
respectively. 
\item{} Reexpress the matrix $M_{\t}$ as a matrix $M_P$. 
\item{} Since $\phi_0$ is cyclic, its canonically conjugate momentum
$P_0$ is conserved. Treat $P_0$ as a constant and form linear
combinations $P_{\pm}'=P_{\pm}+\alpha_{\pm} P_0$ such that the
off-diagonal terms containing $P_0$ are eliminated. A diagonal term
multiplying $P_0^2$ remains. 
\item{} Now diagonalize the lower $2\times2$ block of 
$M_P$. The two eigenvalues of $M_P$ are the squares of the energies of the
normal modes of the hamiltonian.  The third normal mode corresponds to a
uniform rotation of all the spins around the $z$-axis coupled to a change
in $S_z$, and its energy is always zero regardless of the coefficient of $P_0^2$. 
\end{itemize}

The above is true even if there is no ground state degeneracy.  If there
is ground state degeneracy in the $\t$ subspace the matrix $M_P$ has a
null eigenvector $v_0$. For generic $\ta$, $P_0$ has nonzero overlap
with $v_0$. In this case the above procedure always produces a zero
coefficient for $P_0^2$. Thus there is still only one mode of zero
energy, the $S_z$ mode. Another way of seeing this is to recognize that
as long as $P_0$ and $v_0$ have nonzero overlap, one can always rescale
$v_0$ so that it becomes canonically conjugate to $\phi_0$. One then has
to modify $\psi_{\pm}$ to keep them independent of $v_0$ in the Poisson
Bracket sense.

However, if $P_0$ and $v_0$ have zero overlap, the null vector must lie
in the subspace of $P_{\pm}$. This means one of the eigenvalues of the
lower $2\times2$ block of $M_P$ must be zero, implying that there is
{\it another} zero mode of the hamiltonian, apart from the $S_z$
mode. The condition for $P_0$ and $v_0$ to have zero overlap is
identical to ${\partial S_z\over\partial \ta}=0$ along the degeneracy
direction, which is the same as extremizing $S_z$. 

In brief, if ${\partial S_z\over\partial \ta}\ne 0$, there is only one zero
mode, whereas if ${\partial S_z\over\partial \ta}=0$, there are two zero
modes at $\vk=0$.  A rigorous derivation is supplied in the Appendix.

We have done explicit calculations to verify all these statements for
the triangular lattice. If one
computes the frequencies of the $\vk=0$ modes at a non-optimal $\ta$, one 
finds one mode of zero energy (the $S^z$ mode), and two modes of non-zero
energy, neither of which is exactly along the degeneracy
direction. However, at precisely the optimal points, there appear {\it two}
modes with zero energy, one of which is the $S^z$ mode and the other
exactly the degeneracy mode. The third mode still has nonzero energy.

In particular, the argument makes no assumptions about the interactions
other than that they should conserve total $S^z$.  So this result should
hold even for site dilution or longer range interactions.  Of course,
for site dilution, one should focus on a particular realization of
randomness and look at $S^z$ over the degeneracy subspace.  In general,
our argument is that {\it the search for the true ground state can be
  restricted to the points on the degeneracy submanifold where all
  conserved quantities commute in the Poisson Bracket sense with the
  generators of motion along the degeneracy submanifold}.

This criterion can fail if the $\vk\ne0$ modes do not follow the
behavior of the $\vk=0$ modes. Also, if there are more conserved
quantities than degeneracy directions, not all of them may be extremized
at the true ground state. 

With $\ta$ and $\tb$ taking on the value quoted above, we find that $\tc$
is at the extreme edge of its allowed range:
\beq
\cos\tc^*=-{1\over 1-\Delta}\sqrt{{1-2\Delta\over 1-\Delta^2}}
\eeq

\bigskip
\subsection{Spin-Wave Dispersions}
\medskip

\subsubsection{Triangular lattice}

The BZ of the triangular lattice is shown in figure 7, with  the lattice
and reciprocal lattice vectors being
\beqar
\ve_1&=&a(1,0)\cr
\ve_2&=&a(\half,\rtot)\cr
\vG_1&=&{4\pi\over a\sqrt{3}}(\rtot,-\half)\cr
\vG_2&=&{4\pi\over a\sqrt{3}}(0,1)
\eeqar
where $a$ is the lattice spacing. 

As usual, we first concentrate on $h=0$.  Figure 8 shows the spin-wave
dispersion for $\Delta > \half$. Since there is only one sublattice, there is
only one mode. However, plotting it on the reduced BZ of the sublattice
problem forces us to fold it back and represent it as three modes. In
this scheme, there is one gapless mode, which is the Goldstone mode
coresponding to the density fluctuations of the bosons. However, as $\Delta$
approaches $\half$, a ``roton'' minimum develops, precisely at the
wave-vectors corresponding to the sublattice structure, as demonstrated
in figure 9.  At exactly $\Delta=\half$, this becomes a gapless mode,
heralding the transition to the supersolid phase.

Now consider the dispersion at $h=0$ in the SS phase, which is shown in
figure 10.  There are two gapless modes within SWT.  One is the standard
density fluctuation, whereas the other corresponds to the degeneracy
mode. The degeneracy mode will be shown to acquire a gap to higher order
in $1/S$ in the next subsection \cite{shender1,stevek}.  There is a third
``optical'' mode to complete the count of the sublattice degrees of
freedom.  The energy scale of the two low-lying modes is $\Delta$, while the optical
mode has an energy scale of 1.  As we move towards $\Delta = 0$ the two low-lying
modes get softer, until they become completely flat at $\Delta=0$. 

However, when we turn on a field the degeneracy mode becomes gapped, as
shown in figure 11.  Let us now investigate the modes at nonzero field
in the other phases, in particular at the transitions.
The SWT for the canted spin phase can be analysed analytically, since there
is only one sublattice. We find that the SW dispersion is
\beqar
\o(\vk)&=&zS\sqrt{\Delta(1-\g_{\vk})(\Delta+(1-{h^2\over36(1+\Delta)})\g_{\vk})}\cr
\g_{\vk}&=&{1\over z}{\sum_\vdelta}'e^{-i\vk\cdot\vdelta}
\eeqar
where $z=6$ for the triangular lattice, and the prime on the sum restricts
$\vdelta$ to nearest neighbor vectors.
We show an example for $\Delta=0.25,\ h=6.0 $ in figure 12.  We have 
$-\half\le\g_{\vk}\le 1$, which leads to the linear instability
of the SF phase, which is denoted $h_{\rmi 1}$ in equation(11). 
Note that the mean field energies of the various phases lead to
first order instabilities which make the linear instability irrelevant
except at $\Delta=0$ and $\Delta=\half$. 

We next turn to the transition line between the supersolid and \Neel
phases, denoted by $h_{\rmc 1}(\Delta)$, a second order line..
Figure 13 shows the SW dispersion at the transition for $\Delta=0.25$,
$h=0.9738$.  The soft density mode is now at $\vk=0$ and has a quadratic dispersion
instead of the usual linear one.  As we increase $h$ and enter the \Neel
phase, the density mode becomes gapped, as figure 14 shows. The total
energy (including spin wave corrections) in the \Neel phase is independent
of $h$, which demonstrates its incompressibility, and hence the exact
filling of $\twothirds$ throughout this phase.

\subsubsection{\kag Lattice $\vq=0$}
The \kag lattice is a triangular Bravais lattice with a three element basis.
If $a$ is the nearest neighbor separation, then the Bravais lattice constant
is ${\tilde a}=2a$.  The spin wave theory dispersions are quite similar to those
of the triangular lattice, with the modes being softer (because of the lower
coordination).  Note that while the Bravais lattice is triangular, the symmetry
is reduced (to $C_{2v}$) and the dispersion curves do not have zero slope
at the zone edge (the X point).  Figures
15, 16, and 17 show the dispersions in the superfluid, the supersolid, the
\Neel state, respectively.  A noteworthy feature is the presence of a zero
energy mode at $h=0$ along the $\Gamma M$ direction in the supersolid phase.
This mode disperses and has nonzero energy except along the $\Gamma M$ direction.

\subsubsection{\kag 9 Sublattice Structure}

The \roott structure on the \kag lattice is described by a triangular
Bravais lattice of lattice constant ${\tilde a}=2\sqrt{3}\,a$ with a nine
element basis.  The elementary lattice vectors are shown in figure 4.
The spin-wave dispersions are unremarkable except for a flat mode whose
energy vanishes as $\sqrt{h}$ in the small $h$ limit.   This is conected
to the degeneracy mode, which is local up to harmonic order, and will be
discussed in the next subsection. All the other features are quite similar
to those of the triangular lattice as shown by figures 18, 19, and 20. 
 
\bigskip
\subsection{Gap of the Degeneracy Mode}
\medskip

The degeneracy mode appears gapless in SWT (and in fact appears flat in
the \roott structure on the \kag lattice), but acquires a gap to higher
order in the SW expansion\cite{shender1,stevek}. We will first treat the
triangular lattice, and go on to the case of the \kag lattice.

\subsubsection{Triangular lattice}

Let us concentrate on the $\vk=0$ degeneracy mode of the triangular
lattice.  The easiest way to derive the gap is in the spin coherent state
path integral language\cite{assa-book} introduced in the previous
section. 

Now imagine separating the path integral by first integrating all the
modes except the $\vk=0$ degeneracy mode (call the angles $\ta^0$ and
$\phi^0$). 
\beq
{\cal Z}=\int D\ta^0 D\phi^0 \exp-\int d\tau
\left(iS\cos\ta^0{\pz\phi^0\over\pz\tau}
+\half S^2 K(\phi^0)^2\right) \int\prod_{\vk\neq 0} D[\theta_\vk,\phi_\vk]
e^{-\cA'[\theta,\phi]}
\eeq
where $\cA'$ does not include the explicitly written terms involving $\phi_0$ in the
preceding factor.  Since the classical ground state energy does not depend on $\ta^0$,
only the Berry phase term and a ``potential'' energy for $\phi^0$ appear there (expanded
to lowest leading order).  If one neglects the integration over the remainder of the
modes, the degeneracy mode appears to have a potential energy but no
kinetic energy. 
This is analogous to a particle of infinite mass, and from the simple harmonic oscillator
formula $\o=\sqrt{K/M}$, the oscillator energy is zero. This is why the mode appears
gapless. 

However, as seen from figure 2, the integration over the rest of the
modes (to leading order in the spin wave expansion) creates an effective potential
which has a dependence on $\ta^0$, rendering the mass $M$ finite and the
mode gapped. Since the effective potential is produced by SWT, it will
be of order $S$, as opposed to the order $S^2$ potentials generated
classically for the other modes. Thus the gap is order $\sqrt{S}$,
compared to the energies of order $S$ seen in SWT.

Now for the specific details. We concentrate on the neighborhood of
$\ta^*$, the optimal value of $\ta^0$, and write $\ta^0=\ta^*+\delta\ta$. It
is easy to see that the $\phi^0$ that couples to this is
$\phi^0=\fa(\vk=0)-\fb(\vk=0)$. The eigenvector corresponding to this mode is
$(\fa,\fb,\fc)=(\half\phi^0,-\half\phi^0,0)$.  For the triangular lattice the
classical ``potential'' energy of a $\phi^0$ deformation yields
\beq
K=3\D\sin^2\ta^*\left(1+{\sin\tb^*\over2\sin\ta^*}\right)
\eeq
Choosing the  particular value $\D=0.25 $ for illustration, we fit the curve in
figure 2 near its minimum to obtain the term in the effective action 
$\half S C(\delta\ta)^2$, for which we obtain $C\simeq 6$.  Of course, a ``potential''
term will also be generated by the integration of the rest of the modes,
but since it is order $S$, it can be neglected compared to the order
$S^2$ term already present classically.  We will choose the
``coordinate'' as $Q=S\sin\ta^* \phi^0$. The Berry phase term now looks
like $S\sin\ta^*\d\ta(\pz\phi^0/\pz\tau)$, which enables us to identify
$P\equiv\delta\ta$ as the conjugate momentum.  We now write the action as
\beq
S=\int d\tau \left(iP{\pz Q\over\pz\tau}
+\half S C P^2+\half {\tilde K} Q^2 \right)
\eeq
where ${\tilde K}=3\D(1+\half{\sin\tc^*\over \sin\ta^*})$. 
We can now find the gap as the harmonic frequency of this oscillator:
\beq
\o_0=\sqrt{SC{\tilde K}}
\eeq
The numerical value is $\o_0\approx 2.3 \sqrt{S}$. 
The \kag lattice  structure introduces new considerations, which we now address.

\subsubsection{\kag Lattice}

The important difference that occurs for the \kag lattice is that the
effective action for the $\ta$ has a linear cusp at the minimum instead
of a quadratic minimum, as shown in figures 21 and 22 for the $\vq=0$ and
\roott structures, respectively.  This kind of minimum has previously been
seen for the \kag lattice in the fully antiferromagnetic Heisenberg
case \cite{vandelft,ritchey1}.  We assume that the degeneracy mode is
non-dispersing, which, at spin-wave order, is approximately true for the
$\vq=0$ structure, and exactly true for the \roott structure.  This
implies that the modes are local, and we assume this to be true even
after quantum fluctuations have lifted them from zero energy. We can now
interchange the roles of $\theta$ and $\phi$ as $P$ and $Q$ and consider the
quantum mechanics of the Hamiltonian
\beq
{P^2\over2M}+\lambda |Q|
\eeq
where $\lambda$ is of order $S$. This leads to a gap in the degeneracy mode
of order $S^{{2\over3}}$. A lifting of the degeneracy mode of order
$S^{{2\over3}}$ has previously been seen in \cite{chubukov-prl,asakawa2} in
the Heisenberg case. 

Since the \kag \roott structure is always lower in energy, we will
concentrate on it in the following, leaving a detailed analysis of the
$\vq=0$ structure to a future publication. 
The degeneracy mode which was merely gapless at $\vk=0$ on the
triangular lattice, but of nonzero energy at every other $\vk$, becomes
completely flat at zero energy on the \kag lattice with the \roott
structure.  The reason is fairly straightforward.  Consider an ABABAB
hexagon in figure 4b, isolated from the rest of the lattice by a ring of
$C$ sites.  For the optimal angles it is easy to deduce that if
$\ta=\ta^*+\delta\t$, then $\tb=\ta^*-\delta\t$ and that $\tc$ changes only to
order $\delta\t^2$. Any coupling between the $\delta\ta(\vr)$ of neighboring
ABABAB hexagons must be mediated by the bordering $\tc$, and must
therefore be third order or higher in $\delta\ta(\vr)$. This means that to
quadratic order the $\ta$ fluctuations of the hexagon do not interact
with the rest of the lattice. It is these local excitations that produce
the flat mode.  Note that even classically, the energy of this distortion
is not zero if higher orders in $\delta\ta$ are included. This is analogous
to the flat mode in the $\vq=0$ structure of the isotropic \kag
antiferromagnet, where once again, the mode is flat only to harmonic
order\cite{harris}.

Consider this degeneracy mode in the presence of a small field. It is
easy to show that the field energy (per nine site unit cell) near the
optimal angles is
\beq
\d E_{\rm field}=\thalf h S^2 {(2-\Delta)^2\over (1-\Delta)}\cos\ta(\delta\theta)^2\ .
\eeq
We do not need to consider the bond energies of the spins to this order
for the following reasons:

\begin{itemize}
\item{} One can decompose the deviation from the optimal $\ta$, $\tb$,
 $\tc$ into $\delta\tb=\delta\t_{\ssrm BA}+\delta\tb'$, and
 $\delta\tc=\delta\t_{\ssrm CA}+\delta\tc'$,
  where the first part represents the change in $\tb$ and $\tc$ along the
  degeneracy direction, due to the change in $\ta$, and the primed part
  corresponds to a change orthogonal to the degeneracy direction.
 
\item{} The optimal $\delta\tb'$, $\delta\tc'$ in the presence of the field are
  of order $h$. This is because the bond energy is quadratic in $\delta\tb'$,
  $\delta\tc'$ while the field energy is linear in these quantities.  Therefore,
  the energety difference due to the bonds will be of order $h^2$, which can
  be neglected in comparison to the order $h$ field energy at small
  fields.

\item{} There is no bond energy associated with a change along $\delta\ta$. 
\end{itemize}

The local $\phi$ mode conjugate to this $\ta$ mode is
$(\half\delta\phi,-\half\delta\phi,0)$.  The ``potential'' energy (per
nine-site unit cell) of this mode can be calculated from the classical
Hamiltonian exactly as in the triangular lattice case, and is
\beq
U=\thalf\ S^2\Delta\phi^2\sin^2\ta^*\left(2+{\sin\tc^*\over\sin\ta^*}\right)
\eeq
The Berry phase term for this unit cell,\ $ 3S\sin\ta^*\delta\theta(\pz\phi^0/\pz\tau)$,
allows us to identify $Q=3S\sin\ta^*\phi^0$ as the coordinate, and $P=\delta\theta$
as its conjugate momentum.  Then
\beqar
\o(S,h,\Delta)&=&{\tilde{\o}}(\Delta)S\sqrt{h}\cr
{\tilde{\o}}(\Delta)&=&\sqrt{\left(2+{\sin\tc^*\over\sin\ta^*}\right)
{(2-\Delta)^2\over(1-\Delta)}\Delta\cos\ta^*}
\eeqar
This gap goes to zero as $\sqrt{h}$ in the limit $h\tend 0$, and the mode becomes
gapless in addition to being flat. 

So far we have used the field to stiffen the degeneracy mode. Now
consider the effect of quantum fluctuations at $h=0$. 
We can find the effective potential to lowest leading order by computing
the ground state energy for various $\ta$ in SWT. This is plotted for $\D=0.25$  in
figure 22. We assume that the mode will remain dispersionless, and therefore local,
even when lifted from zero energy by quantum fluctuations.  This assumption allows
us to extract an effective Hamiltonian for each local degeneracy mode,
which has been written down in equation(35), 
where we identify the {\it momentum} as $P=3S\sin\ta^*\phi^0$ and
the conjugate {\it coordinate} as $Q=\delta\theta$, and where the numerical
value of $\lambda$ is obtained from figure (for $\Delta=0.25$):
\beqar
M^{-1}&=& 0.1944\cr
\lambda&=&0.025\, S
\eeqar
We use a Gaussian trial wave function to obtain the
approximate ground state energy of this Hamiltonian, yielding
\beq
\o_0\approx\left({27\lambda^2M^{-1}\over128\pi}\right)^{{1\over3}}\approx
0.02\,S^{{2\over3}}
\eeq
Of course, we expect this mode to disperse, but the calculation of the
dispersion\cite{chubukov-prl,asakawa2} is much harder than the one presented
here, and will be pursued in future work.

\bigskip
\section{Effect of Quantum Fluctuations on Order Parameters}
\bigskip

One of the motivations for this work has been to see if quantum
fluctuations can disorder the system, creating a spin liquid, which
would correspond to a ordinary liquid (with nonzero viscosity) for the
bosons. To leading order in SWT, one can compute the average values of
the spins as
\beq
\langle \vS_i\rangle = (\langle a\yd_i a\nd_i \rangle -S){\hat\Omega}_\MF\ .
\eeq
The calculation sketched out in section II also produces the explicit
Bogoliubov transformation, which can then be used to find the
expectation values of bilinears. More explicitly, in terms of the matrix $T$
which implements the transformation 
\beq
\langle e\yd_i(\vk)\,e\nd_j(\vk)\rangle=\sum_{\alpha=K+1}^{2K} T\yd_{\alpha i}(\vk)\,
T\nd_{j \alpha}(\vk)\ ,
\eeq
where it is understood that $i$ and $j$ run from 1 to $K$ (the number of
sublattices), while $\alpha$ runs from $K+1$ to $2K$, the rank of $T$.
Let us now turn to the different cases. 

\bigskip
\subsection{Triangular Lattice}
\medskip

Figure 3 shows a plot of the classical and quantum-corrected total $S^z$
and total $S^x$, for $h=0$.  It is clear that the quantum-corrected
value of the magnetization is very close to zero, independent of
$\Delta$.  Exact diagonalizations of finite clusters for the anisotropic
antiferromagnetic case $\Delta<0$\cite{pt+exact-tri} suggest strongly
that this is an exact statement.  That is, the exact ground state at
$h=0$ has zero longitudinal magnetization $M_z$.  An interesting fact
about the triangular lattice allows us to map the $\Delta>0$ problem
(our model) to the $\Delta<0$ model, which is the anisotropic
antiferromagnet solved in SWT by KMFF.  For very small $\Delta$ one can
work to linear order in $\Delta$, which means that one is working within
the set of antiferromagnetic Ising model ground states.  For
nearest-neighbor spin flips, the Marshall sign property is obeyed by the
wave-functions, arising from a partition of the set of ground states
into disjoint even and odd states \cite{blote}.  This means that for
very small $\Delta$ there is evidence that the boson system would be
exactly half-filled in the true ground state.  However, to any higher
order in $\Delta$, or numerically for $\Delta$ not so small, the two
models have no simple relationship with each other since the triangular
lattice is not bipartite.

Although total $S^x$ decreases as $\Delta$ decreases, it seems that
superfluid order persists all the way, vanishing only when $\Delta=0$.  In
figure 23 we plot the quantum fluctuation corrections to the magnitudes of
the spins on the three sublattices (since $\tb=\tc$ we plot only two
values).  As $\Delta\tend 0$ the A sublattice remains stiff, while the B
and C sublattice spins get reduced to about half their classical value
(for $S=\half$).  Thus, supersolid order on the triangular lattice
survives quantum fluctuations at the spin wave level.

Let us now turn to figure 24, which shows the classical and quantum
order parameters as a function of $\D$  for $h=3$. For $\D<\half$ it is
clear that though there are quantum fluctuation corrections to each of
the spins, the total $S_z$ is not corrected, thus leaving the filling
exactly ${2\over3}$. For $\D>\half$, there are fluctuation corrections
which do not affect the nature of the phase. 

Once again, we will concentrate on the lower energy \roott structure on
the \kag lattice, leaving an analysis of the $\vq=0$ structure to future
work.

\bigskip
\subsection{\kag 9 Sublattice Structure}
\medskip

Figure 5 presents the na{\"\i}ve results for the occupation numbers of the
different sublattices in SWT for the \roott structure at $\Delta=0.25$ as a
function of $h$. Notice that as $h\tend0$ the fluctuation corrections
diverge.  This is a consequence of the flat mode, whose energy goes to
zero as $h\tend0$.  In reality, as we have seen in the previous section,
the energy flat mode will be lifted in higher orders of SWT.  While a
quantitatively accurate analysis requires the computation of the
dispersion of this degeneracy mode, we can make a rough estimate of the
fluctuation corrections to the sublattice spins by assuming that the
mode remains flat at the value $\o_1$ calculated in Section II.C.2. 

We need to determine the reduction in the magnitude of each sublattice
spin due to quantum fluctuations. In terms of the equilibrium position
$\t^*$ and the deviations from equilibrium $\d\t$ and $\d\phi$ we can
write the magnitude of the spin as

\beq
|\langle\vS\rangle|=S\left(1-\half(\delta\theta)^2-\half\sin^2\theta^*\,\phi^2\right)=
\left(1- {P^2\over 2} -{Q^2\over18S^2}\right) 
\eeq
where we have once again used $P=3S\sin\ta^*\phi^0$ and $Q=\ta$, and
assumed a non-dispersing degeneracy mode. 

We know from the harmonic oscillator that 
\beqar
\langle P^2\rangle_0&=&{\o\over2M^{-1}}\cr
\langle Q^2\rangle_0&=&{\o\over2K}
\eeqar
and in the case $h\ne 0$, we find
\beq
|\langle\vS\rangle|=S-{1\over12S\sqrt{h}}\left(\sqrt{{\D(2+{\sin\tc^*\over\sin\ta^*})\over
\cos\ta^*{(2-\D)^2\over1-\D}}}+{h\over2}\sqrt{{\cos\ta^*{(2-\D)^2\over1-\D}
\over\D(2+{\sin\ta^*\over\sin\ta^*})}}\right)
\eeq

Clearly there is a divergence as $h\tend0$, which is seen in figure 5. 
Of course, there will also be a contribution independent of $h$ due to
the other modes not associated with the degeneracy.

Now consider the case $h=0$.  As shown in the previous section, the flat
mode will be lifted from zero energy by quantum fluctuations and acquire
an energy of order $S^{2\over3}$.  We will assume that the mode does not
disperse  (even though it will, weakly) in order to estimate the quantum
fluctuations in the sublattice spins.  We will also assume that none of
the matrix elements change substantially.  Thus the energy of the mode is
the only significant factor.  This means that we can
estimate the contribution to the fluctuation correction to the spins by
applying a field $h_Q$ such that the field-induced energy is the same as the
energy induced by quantum fluctuations:
\beq
h_Q\approx {\lambda^2\over{\tilde{\o}}^2}\, S^{-{2\over3}}\ .
\eeq

Numerically, we find for $\Delta=0.25$ that $h_Q=0.00057$.  This implies
boson occupations of $n_{\ssrm A}=n_{\ssrm B}=1.66$, and $n_{\ssrm C}=0.04$.
Since the boson
occupations are larger than the value $S=\half$, we find the
fluctuations to be larger than the mean field value of the spin.  Thus,
despite the lifting of the flat mode due to quantum fluctuations, we find
that fluctuations may be large enough to disorder the mean field
ordering on the A and B sublattices.  However, it is not certain that
large fluctuations imply disorder.  Furthermore, it is difficult to
understand how the C-sublattice could remain stiff and ordered if the
other two lattices are disordered. Strictly speaking, what this result
tells us is that we have reached the limitations of spin-wave
theory. 

\bigskip
\section{Conclusions and Open Questions}
\medskip

We have found a broad region of parameter space where supersolid order
is stable to leading order in mean field theory.  The inclusion of
spin wave corrections modifies this picture differently for the
triangular and \kag lattices.  The supersolid is quite robust on the
triangular lattice.  Our interpretation of the supersolid corresponds
with the ideas of Andreev and Lifshitz\cite{andreev}, with hopping vacancies
undergoing Bose condensation.  The role of lattice frustration is to
prevent all the particles from condensing into a solid. 

We have also come across a general (though nonrigorous) argument which
limits the search for the true ground state in a system with ground
state degeneracy: Look for the points at which global conserved
quantities commute in the Poisson Bracket sense with the generators of
motion on the degeneracy subspace.

On the \kag lattice we find the \roott structure to be more stable than
the $\vq=0$ structure at all parameter values.  Fluctuations seem much
stronger here, and may even be able to destroy the long-range order
assumed in mean field theory.  This is a fertile region for numerical and
experimental work.

To what experimental systems might these considerations be applicable?
One might be able to construct an array of Josephson
junctions\cite{geerligs}  which
satisfy the conditions necessary for the existence of a supersolid.  This
means that the charging energy of each grain should be very high, and
the nearest-neighbor charging energy should be higher than the Josephson
coupling between the neighboring grains.  Furthermore, only pair hopping
should be relevant, which implies temperatures low compared to the bulk
superconducting $T_\rmc$. Of course, since this is a two-dimensional
system, the Bose condensate disappears for $T\ne 0$, but power-law ODLRO
is expected to remain. 

Another experimental system to which these results might be relevant is
\hef on graphite.  A variety of orderings and transitions are known to
occur as a function of temperature and coverage\cite{bretz}.  Also,
steps in the superfluid density have been seen as a function of coverage
(for multiple layers) for \hef on graphite\cite{crowell}, and have been
interpreted as resulting from correlation effects\cite{zim-crow}.

As it stands, this work is {\it not} applicable to the question of
supersolidity in \hef on a smooth substrate.  In order to approach
the continuum one would have to consider very low densities on the
lattice, as well as long-range interactions (see ref\cite{pomeau} for an
example of a continuum approach). 

We close with a number of important open questions. 

SWT seems sufficient for the triangular lattice, but not for the \kag
lattice.  A formalism that can consider ordered and disordered states in
a unified manner is necessary, perhaps a variant of the large-$N$
approaches\cite{sun,subir-spn}.  Secondly, even for the triangular
lattice, the question of vortices in the ground state is open. The
spin-wave ground state has vortices, but they occur in nearest-neighbor
pairs.  A fruitful way to incorporate vortices might be to consider the
effective theory of the gapless modes only, and include their
interaction with vortices in a semiclassical manner.

The original picture of ref\cite{fwgf} has been confirmed by a
strong-coupling expansion\cite{jfhm}. Strong-couplings expansions and
variational approaches\cite{rokhsar,krauth} are complimentary to the
spin-wave expansion, and it would be interesting to investigate our
model by these approaches.

The effects of nonzero temperature on systems with ground state
degeneracy can be very nontrivial.  In particular, quantum and thermal
ground state selection effects may compete to produce a sequence of
phases and transitions as the temperature is raised \cite{sheng}.
Clearly, thermal effects have to be elucidated before our results can be
directly applied to an experimental situation.

Finally, it is intriguing to harken back to the question raised by
Anderson and Fazekas\cite{fazekas-pwa} and ask whether it is possible
for a liquid state to be stable at zero temperature.

\section{Acknowledgements}
It is a pleasure to thank John Clarke, Subir Sachdev, Steve Kivelson,
Dung-hai Lee, Shoucheng Zhang, Zlatko Tesanovic, and Dan Rokhsar for
valuable discussions.  We are especially grateful to Chris Henley for
detailed comments on draft versions and enlightening discussions. This
research was supported in part by NSF grant \# DMR-9311949 (GM).
GM and DPA are also grateful to the hospitality of the Technion at Haifa,
where this work was initiated.

\bigskip
\subsection{Appendix: Continuous Degeneracy not Arising from a Symmetry}
\medskip

\def\ij{_{ij}}
\def\mn{{\mu\nu}}
\def\RT{R^{\rm t}}
\def\pT{p^{\rm t}}
\def\qT{q^{\rm t}}
\def\NN{{N\times N}}
\def\hbot{\frac{\hbar}{2}}
\def\Vx{{\vec x}}
\def\Vp{{\vec p}}
\def\VP{{\vec P}}
\def\Vq{{\vec q}}
\def\VQ{{\vec Q}}
\def\VX{{\vec X}}
\def\cD{{\cal D}}
\def\ket#1{{\,|\,#1\,\rangle\,}}
\def\bra#1{{\,\langle\,#1\,|\,}}
\def\braket#1#2{{\,\langle\,#1\,|\,#2\,\rangle\,}}
\def\expect#1#2#3{{\,\langle\,#1\,|\,#2\,|\,#3\,\rangle\,}}

Consider a Hamiltonian function $\cH(\Vx)$, where
\beq
\Vx=(p\nd_1,p\nd_2,\ldots,p\nd_N,q\nd_1,q\nd_2,\ldots,q\nd_N)
\eeq
is the vector of (dimensionless) coordinates and momenta.  What does it mean
that the ground state is continuously degenerate?  Clearly, there must exist
a one-parameter family of ground states, \ie
\beq
\left({\pz\cH\over\pz x_\mu}\right)_{\VX(\lambda)}=0
\eeq
for all $\mu=1,\ldots,2N$, where $\lambda$ parametrizes the degeneracy
submanifold, a curve in phase space defined by $\VX(\lambda)$.  The degeneracy
submanifold is one-dimensional in this example.  Differentiating with respect
to $\lambda$ gives
\beq
\sum_{\nu=1}^{2N}\left({\pz^2\cH\over\pz x_\mu\, \pz x_\nu}\right)_{\VX(\lambda)}
{\pz X_\nu(\lambda)\over\pz\lambda}=0\ ,
\eeq
\ie\ the vector $\pz X_\nu/\pz \lambda$ is a null eigenvector of the
Hessian matrix evaluated at the point $\VX(\lambda)$.

As an example, consider the function
\beq
\cH(r,\phi)=\fourth (r^2-1)^2\, (a+b\cos\phi)
\eeq
whose minimum is at $r=1$ independent of $\phi$.  Note that $\phi$ is not
cyclic in $\cH$, and its conjugate momentum (say $r^2$, by definition) is
not conserved as Noether's theorem does not apply.
Nonetheless, there is a one-parameter family of
ground states, parametrized by $\phi$.  The Hessian matrix, evaluated at $r=1$, is
\beq
H=\pmatrix{2(a+b\cos\phi)&&0\cr&&\cr 0&&0\cr}
\eeq
and the vector ${\pz\over\pz\phi}\,\pmatrix{r^2 \cr\phi\cr}=
\pmatrix{0\cr 1\cr}$ is a null eigenvector.

Now consider the Gaussian fluctuations about the ground state.
We are interested in the partition function $Z=\int\cD[p,q]\,e^{-\cA}\>$,
which is obtained from the action functional
\beq
\cA=\int_0^{\beta\hbar}\!\!d\tau\,\left(i\hbar\sum_i q_i {\pz p_i\over\pz \tau}
+\cH(\Vp,\Vq)\right)
\eeq
where $i=1,\ldots,N$.  We expand about one of the ground states, writing
$p_i=P_i+\delta p_i$, $q_i=Q_i+\delta q_i$, and, using summation convention,
\beq
\cH=\hbot\, \delta p_i\, T\nd\ij\,\delta p_j+\hbot\,\delta p_i\, R\nd\ij\,
\delta q_j +\hbot\,\delta q_i\,\RT\ij\,\delta p_j +\hbot\,\delta q_i\,
V\nd\ij\,\delta  q_j
\eeq
with
\beq
T\ij={1\over\hbar}\left({\pz^2\cH\over\pz p_i\,\pz p_j}\right)_{\VP,\VQ}\qquad
R\ij={1\over\hbar}\left({\pz^2\cH\over\pz p_i\,\pz q_j}\right)_{\VP,\VQ}\qquad
V\ij={1\over\hbar}\left({\pz^2\cH\over\pz q_i\,\pz q_j}\right)_{\VP,\VQ}
\eeq
and $\RT$ is the transpose of $R$.  Thus, the integrand of $\cA$ is given by
$\hbot \delta x_\mu\, (H_\mn+i J_\mn\pz_\tau)\, \delta x_\nu$, where
\beq
H=\pmatrix{T&&R\cr&&\cr\RT&&V\cr}
\eeq
and
\beq
J=\pmatrix{0\nd_\NN & & 1\nd_\NN\cr & & \cr -1\nd_\NN && 0\nd_\NN \cr}\ .
\eeq
The matrix $H$ is brought to diagonal form by the basis change $S$:
\beq
S\yd H S = \pmatrix{\Omega^+&&0\cr &&\cr 0 &&\Omega^-\cr}
\eeq
where $\Omega^\pm={\rm diag}(\omega^\pm_1,\ldots,\omega^\pm_N)$
is a diagonal matrix, and $S\yd J S=J$ preserves the symplectic structure
(commutation relations).  Note that the diagonalized Hamiltonian
is $\half(\pT\,\Omega^+\, p +\qT\,\Omega^-\, q)$ rather than, say,
$\half (\pT\, p + \qT\, \Omega^2\, q)$.  The two are, of course, related
by a canonical transformation, provided the $\omega^\pm_k$
are nonzero.)  The eigenvectors come in pairs $\ket{\psi^\pm_k}$
which satisfy
\beqar
JH\ket{\psi^+_k}&=&-\omega^+_k\ket{\psi^-_k}\nonumber\cr
JH\ket{\psi^-_k}&=&+\omega^-_k\ket{\psi^+_k}
\eeqar
as well as the orthonormalization condition
\beqar
\expect{\psi^+_k}{J}{\psi^+_l}&=&\expect{\psi^-_k}{J}
{\psi^-_l}=0\nonumber\cr
-\expect{\psi^-_k}{J}{\psi^+_l}&=&\expect{\psi^+_k}{J}
{\psi^-_l}=\delta_{kl}\ ,
\eeqar
where $k,l=1,\ldots,N$.  The $N$ energy eigenvalues are then given
by $\ve_k=\hbar\sqrt{\omega^+_k\,\omega^-_k}$.

Now consider the case where the degeneracy subspace is two-dimensional,
parametrized by the coordinates $(\lambda,\xi)$.  In our spin wave theory,
we have that $q_i=\phi_i$ and $p_i=S(1-\cos\theta_i)$, and the $\lambda$
invariance is realized as some relation among the colatitutes
$\Theta_i(\lambda)$ along the ground state submanifold, while the $\xi$
invariance is simply expressed as $\Phi_i(\xi)=\Phi_i(0)+\xi$.
The matrix $H$ (and hence $JH$) has two known null right eigenvectors -- call
them $\ket{N_1}$ and $\ket{N_2}$ -- corresponding,
respectively, to the $\lambda$ and $\xi$ invariances.  We have that
\beq
\ket{N_1}=\pmatrix{{\pz P_i\over \pz\lambda}\cr \cr {\pz Q_i\over
\pz\lambda}\cr}
=\pmatrix{S\sin\Theta_i\, {\pz\Theta_i\over\pz\lambda}\cr \cr 0\nd_N\cr}
\eeq
and
\beq
\ket{N_2}=\pmatrix{{\pz P_i\over \pz\xi}\cr \cr {\pz Q_i\over \pz\xi}\cr}
=\pmatrix{0\nd_N\cr \cr 1\nd_N\cr}\ .
\eeq
Note that the $\ket{N_2}$ eigenvector contains a 1 in each of its lower $N$
entries.  When we compute the overlap $\expect{N_2}{J}{N_1}$, we obtain
\beq
\expect{N_2}{J}{N_1}=-\sum_{i=1}^N S\sin\Theta_i{\pz\Theta_i\over\pz\lambda}
={\pz M^z\over\pz\lambda}\ ,
\eeq
where $M^z$ is the total $\zhat$ component of the magnetization:
$M^z\equiv\sum_{i=1}^N S\cos\Theta_i$.  Unless the magnetization $M_z$ is
extremized with respect to $\lambda$, there is a finite overlap, and by
appropriate normalization one can choose $\ket{N_1}$ and $\ket{N_2}$ to be
a $\ket{\psi^\pm}$ pair.  When the magnetization is extremized, however,
the overlap is zero which means that these two null right eigenvectors belong
to separate pairs, and there are at least {\it two} zero modes.

Therefore, {\it the two modes are independent only when $\pz M^z/\pz\lambda=0$,
that is to say, when the magnetization $M^z$ is extremized.}

In the general case both the conserved quantity and the generator of
motion along the degeneracy subspace may have components along both $p$-
and $q$-subspaces. It is then easy to see that the two vectors will have
zero overlap only if the conserved quantity commutes (in the Poisson
Bracket sense) with the generator of motion along the degeneracy
subspace.

\begin{figure}
\caption{Mean-field phase diagram for the triangular lattice. The Kagom\'e
  lattice phase diagram differs only by a rescaling of $h$. Heavy lines
  denote first order transitions, light lines second-order transitions,
  and dashed lines denote linear instabilities.}
\label{fig1}
\end{figure}

\begin{figure}
\caption{Ground state energy of the triangular lattice at $\D=0.25$ as a
  function of $\ta$. The minimum is quadratic. }
\label{fig2}
\end{figure}

\begin{figure}
\caption{Classical and quantum magnetizations for the triangular
  lattice, $S_z$ and $S_x$ per site, as a function of $\D$ at the true
  ground state. The full quantum-corrected $S_z$ is always close to
  zero, while the quantum-corrected $S_x$ is always nonzero.}
\label{fig3}
\end{figure}

\begin{figure}
\caption{The two long range ordered structures on the Kagom\'e lattice.
  (a) the $q=0$ structure and (b) the R3 structure.}
\label{fig4}
\end{figure}

\begin{figure}
\caption{Classical and quantum-corrected values of the magnetizations
  $S_z$ and $S_x$ per site as a function of $h$ at $\D=0.25$ for the
  R3 structure on the Kagom\'e lattice. Notice the divergence as
  $h\rightarrow0$. This is an artifact of SWT.}
\label{fig5}
\end{figure}

\begin{figure}
\caption{Optimal $\ta=\tb$ and $\tc$ as a function of $h$ for
  $\D=0.25$. Note the continuous approach to collinearity at the
  supersolid-solid transition, and the discontinuous change of the
  angles at the solid-superfluid transition.}
\label{fig6}
\end{figure}

\begin{figure}
\caption{The Brillouin Zone of the triangular lattice and the reduced
  zone for the sublattice structure. }  
\label{fig7}
\end{figure}

\begin{figure}
\caption{Triangular lattice SW dispersion for $\D=0.75$, $h=0$.}  
\label{fig8}
\end{figure}

\begin{figure}
\caption{Triangular lattice SW dispersion for $\D=0.51$, $h=0$. 
  Note the quadratic minimum which is nearly gapless. This is the linear
  instability that leads to the SS phase.}
\label{fig9}
\end{figure}

\begin{figure}
\caption{Triangular lattice SW dispersion in the SS phase at 
  $\D=0.25$, $h=0$.}
\label{fig10}
\end{figure}

\begin{figure}
\caption{Triangular lattice SW dispersion in the SS phase at $\D=0.25$, 
  $h=0.5$. The degeneracy mode has become gapped}
\label{fig11}
\end{figure}

\begin{figure}
\caption{Triangular lattice SW dispersion in the CS phase at $\D=0.25$, 
  $h=6.0$. Note the single gapless density mode.}
\label{fig12}
\end{figure}

\begin{figure}
\caption{Triangular lattice SW dispersion at the SS-MI transition at $\D=0.25$,
  $h=0.9738$. Note the quadratic dispersion of the gapless density mode.}  
\label{fig13}
\end{figure}

\begin{figure}
\caption{Triangular lattice SW dispersion in the MI phase at $\D=0.25$, $h=1.5$.}  
\label{fig14}
\end{figure}

\begin{figure}
\caption{Kagom\'e $\vq=0$ SW dispersion in the CS phase at $\D=0.25$, $h=4$.}  
\label{fig15}
\end{figure}

\begin{figure}
\caption{Kagom\'e $\vq=0$ SW dispersion in the SS phase at $\D=0.25$,
  $h=0$. Note the flat chharacter of the mode along the $\Gamma M$
  direction.}  
\label{fig16}
\end{figure}

\begin{figure}
\caption{Kagom\'e $\vq=0$ SW dispersion in the MI phase at $\D=0.25$, $h=1$.}  
\label{fig17}
\end{figure}

\begin{figure}
\caption{Kagom\'e R3 SW dispersion in the CS phase at $\D=0.25$, $h=4$. }  
\label{fig18}
\end{figure}

\begin{figure}
\caption{Kagom\'e R3 SW dispersion in the SS phase at $\D=0.25$,
  $h=0$. Note the completely flat zero energy mode.}  
\label{fig19}
\end{figure}

\begin{figure}
\caption{Kagom\'e R3 SW dispersion in the MI phase at $\D=0.25$, $h=1$.}  
\label{fig20}
\end{figure}

\begin{figure}
\caption{Quantum energy correction versus $\ta$ for the Kagom\'e $\vq=0$
  structure. Note the cusp at the minimum.}  
\label{fig21}
\end{figure}

\begin{figure}
\caption{Quantum energy correction versus $\ta$ for the Kagom\'e R3 
  structure. Once again the leading order potential has a linear cusp.}  
\label{fig22}
\end{figure}

\begin{figure}
\caption{Quantum fluctuation corrections to the magnitudes of the spins
  on the $A$ and $C$ sublattices as a function of $\D$ at $h=0$ for the
  triangular lattice.}  
\label{fig23}
\end{figure}

\begin{figure}
\caption{Classical and quantum values of $S_z$ and $S_x$ for the
  triangular lattice at $h=3$ as a function of $\D$. Note that for
  $\D<\half$ we are in the MI phase, and the total magnetization is
  uncorrected from its mean-field value of ${2\over3}$. }
\label{fig24}
\end{figure}


\begin{references}
\bibitem{stevek} We are indebted to Steve Kivelson for making us aware
of the calculation of the gap of the degeneracy mode in ref. \cite{shender1}.
\bibitem{henley-pc} We are indebted to Chris Henley for making us aware
  of weaknesses in our original argument.
\end{references}
\end{document}